\def\aa{{A\&A}}

\def\aj{{AJ}}

\def\apj{{ApJ}}
\def\apjs{{ApJS}}

\def\mnras{{MNRAS}}

\documentclass[cup5b]{caps}
\usepackage{graphicx}
\usepackage{amssymb}
\usepackage{ociwsymp3}   

\HeadText{R. G. Bower and M. L. Balogh}

\begin{document}

\pagenumbering{arabic}

\author[]{RICHARD G. BOWER \& MICHAEL L. BALOGH\\Institute for Computational 
Cosmology, University of Durham, UK}

\chapter{The Difference Between Clusters and \\
         Groups: A Journey from Cluster \\
         Cores to Their Outskirts and Beyond}

\begin{abstract}

In this review, we take the reader on a journey. We start by looking at
the properties of galaxies in the cores of rich clusters. We have
focused on the overall picture: star formation in clusters is
strongly suppressed relative to field galaxies at the same redshift. We
will argue that the increasing activity and blue populations of
clusters with redshift results from a greater level of activity in
field galaxies rather than a change in the transformation imposed by
the cluster environment. With this in mind, we travel out from the
cluster, focusing first on the properties of galaxies in the
outskirts of clusters and then on galaxies in isolated groups.
At low redshift, we are able to efficiently probe these environments
using the Sloan Digital Sky Survey and 2dF redshift surveys. These allow an 
accurate comparison of galaxy star formation rates in different regions.
The current results show a strong suppression of star
formation above a critical threshold in local density.  The threshold
seems similar regardless of the overall mass of the system.  At low
redshift at least, only galaxies in close, isolated pairs have their
star formation rate boosted above the global average. At higher
redshift, work on constructing homogeneous catalogs of galaxies in
groups and in the infall regions of clusters is still at an early
stage. In the final section, we draw these strands together,
summarizing what we can deduce about the mechanisms that transform
star-forming field galaxies into their quiescent cluster counterparts.
We discuss what we can learn about the impact of environment on the
global star formation history of the Universe.

\end{abstract}

\def\Msol{\,M_{\odot}}
\def\Mpc{\,\hbox{Mpc}}
\def\kms{\,\hbox{km}\,\hbox{s}^{-1}}

\section{Introduction}

Let us start with an outline of this review. We will begin by looking at
galaxies in the cores of clusters. We have been observing clusters for many
years. Some milestones are the papers on the morphological differences
between cluster galaxies and the general field (Hubble \& Humason
1931), the discovery of a global morphology-density relation (Oemler
1974; Dressler 1980), and the realization of the importance of the
color-magnitude relation (Sandage \& Visvanathan 1978).  We will
attempt to summarize what we have learned from looking at clusters
since this time. In particular, recent observations now span a wide
range of redshift, allowing us to look directly at how the galaxy
populations evolve.

In the second section, we will
investigate how galaxy star formation rates vary with radius and
local density. In particular, we will focus on the recent results
from the 2dF galaxy redshift survey. The aim here is to understand how
galaxy properties are influenced by their environment. As we will
discuss, it seems that the group environment is critical to the
evolution of galaxies, creating a distinctive threshold.

In the third section, we will review some of the ideas about how this can
all be put together, and how we can hope to use the environmental
studies that many groups are undertaking to build a better
understanding of the evolution of the Universe.

Throughout, this paper will focus on galaxy star formation rates as
the measure of galaxy properties, and will leave aside the whole issue
of galaxy morphology for other reviewers to deal with. Clearly, the
two issues are related since galaxy morphology partly reflects the
strength of H~{\sc ii} regions in the galaxy disk (Sandage 1961), but the
two factors are not uniquely linked. Morphology and star formation may
be influenced differently by different environments (Dressler et al.\
1997; Poggianti et al.\ 1999; McIntosh, Rix, \& Caldwell 2003).
We will also skirt around the important
issue of E+A galaxies (Couch \& Sharples 1987; Dressler \& Gunn 1992; 
Barger et al.\ 1996; Balogh et al. 1999; Poggianti et al.\ 1999) and
star formation that is obscured from view in the optical (Poggianti \& Wu 1999;
Smail et al. 1999; Duc et al.\ 2002; Miller \& Owen 2002).
These are discussed in detail in Poggianti (2003). Wherever
possible we will use H$\alpha$ as the star formation indicator (Kennicutt 
1992), but as we probe to higher redshift, we are forced to use [O~{\sc ii}] 
$\lambda$3727 unless we shift our strategy to infrared spectrographs.

We will also stick to talking about bright galaxies, by which we mean
galaxies brighter than 1~mag fainter than $L_*$. It would need another
complete review if we were to compare the properties of dwarf galaxies
over the same range of environments. A good place to start would be
Drinkwater et al.\ (2001), or the many presentations on cluster dwarfs
at this Symposium. By the same token, we will avoid discussion of the
evolution of the galaxy luminosity function (Barger et al.\ 1998; De~Propris 
et al.\ 1999); this is summarized in Rudnick et al.\ (2003).

To avoid confusion, it is worth laying out exactly what we mean by the terms 
``cluster'' and ``group.'' We will use the term cluster to mean a virialized 
halo with mass greater than $10^{14} \Msol$ and the term group to mean a halo
more massive than about $10^{13} \Msol$ (but less than $10^{14} \Msol$).
If an isolated $L_*$
galaxy has a halo mass of order $10^{12} \Msol$ (Evans \& Wilkinson 2000;
Guzik \& Seljak 2002; Sakamoto, Chiba, \& Beers 2003), then our definition of a
group contains more than five $L_*$ galaxies at the present day. At higher
redshift, the conversion between mass and galaxy numbers is more
complicated since it depends on whether $L_*$ evolves or not. If we
stick to a definition in terms of mass, then at least everything is
clear from a theoretical perspective, and we can make quite definite
predictions about the numbers of such halos, their clustering as a
function of redshift (Press \& Schechter 1974; Jenkins et al. 2001; 
Sheth, Mo, \& Tormen 2001), and how mass accumulates from smaller halos
into large clusters (Bond et al. 1991; Bower 1991; Lacey \& Cole
1993; Mo \& White 2002).

\section{Clusters of Galaxies}

At the outset, its worth reminding ourselves of why we study galaxy
evolution in clusters.
One popular reason is that the cluster is a good laboratory in which
to study galaxy evolution. Another is that it is ``easy'' ---
when we observe the galaxy spectra, we know that most objects will be
in this dense environment and that our observations will be highly
efficient. The same reason allows us to recognize clusters out to very
high redshifts and thus to extend our studies to a very long baseline. But we 
should remember that clusters do have a significant
drawback: they are rare objects. For the standard $\Lambda$CDM cosmology
($\Omega_{\rm m}=0.3$, $\Omega_\Lambda=0.7$, $h=0.7$, $\sigma_8=0.9$),
the space density of $>10^{14} \Msol$ halos is $7\times10^{-5}
h^{-3}\, {\rm Mpc}^{-3}$. Even though such
clusters contain $\sim 100$ $L_*$ galaxies, less than 10\% of the cosmic
galaxy population is found in such objects.

There is an emerging consensus that suggests that the stellar
populations of galaxies in cluster cores are generally old, with most of
the stars formed at $z>2$. Most of these galaxies also have early-type
morphology. It is possible to derive remarkably tight
constraints from looking at colors (Bower, Lucey, \& Ellis 1992; Bower, 
Kodama, \& Terlevich 1998; Gladders et al.\ 1998; van~Dokkum et al.\ 1998),
at the Mg-$\sigma$ relation (G\'uzman et al. 1992),
or at the scatter in the fundamental plane (J{\o}rgensen et al. 1999;
Fritz et al. 2003).
These results rely on the argument that recent star formation would lead
to excessive scatter in these tight relations, unless it was in some way
coordinated, or the color variations due to age were cancelled out by
variations in metal abundance (Faber et al. 1999; Ferreras, Charlot, \&
Silk 1999).
Line-index measurements  generally suggest very old populations
(J{\o}rgensen 1999; Poggianti et al.\ 2001), but these relations tend to
show somewhat more scatter. This has been interpreted as
evidence for the cancellation effects in broad-band colors.

To improve the evidence, one can compare clusters at high redshift. For example,
if we concentrate on the color-magnitude relation, we would expect the
narrow relation seen in local clusters to break down as we approach
the epoch when star formation was prevalent. In fact, we have
discovered that the color-magnitude relation is well established in high-redshift clusters
(Ellis et al. 1997; van~Dokkum et al.\ 1998), and that the line-index
correlations, fundamental plane (Kelson et al.\ 2001), and
Tully-Fisher relation measurements (e.g., Metavier 2003; Ziegler et al.\ 2003;
but see Milvang-Jensen et al.\ 2003) also show little increase
in scatter compared to local clusters. So far, tight relations have been
identified in clusters out to $z=1.27$ (van~Dokkum et al.\ 2000; Barrientos et 
al.\ 2003; van~Dokkum \& Stanford 2003).
The tight relation does eventually seem to break down, and we are not aware
of any strong color-magnitude relation that has been
identified in ``proto-clusters'' at $z>2$.

There is a bias here, however, that should be clearly
recognized . Although we are discovering that clusters at high
redshifts seem also to contain old galaxies, this does not mean
that all galaxies in local clusters must have these old populations. A
large fraction of galaxies that are bound into local clusters would
have been isolated ``field'' galaxies at $z\approx 1$.
An even stronger bias of this type has been termed ``progenitor bias''
by van~Dokkum \& Franx (2001). They point out that if only a subset of the
cluster populations is studied (for example only the galaxies with
early-type morphology), then it is quite easy to arrive at a biased
view. To get the full picture, one needs to study the galaxy
population of the cluster as a whole.

An interesting strategy is therefore to simply measure the star
formation rate in clusters at different epochs. The general consensus
seems to be that there is little star formation (relative to field
galaxies at the same redshift) in virialized cluster cores below $z=1.5$.
For example, Couch et al.'s (2001)
survey of the AC114 cluster found that star formation was suppressed
by an order of magnitude compared to the field.  Similar levels of
suppression are seen in poor clusters (Balogh et al. 2002).  While
these studies find some exciting objects (see Finn \& Zaritsky 2003
for further examples), the general trend is for the star formation
rate to be strongly suppressed relative to the field at the same
redshift. infrared measurements (Duc et al.\ 2002) and radio measurements
(Miller 2003; Morrison \& Owen 2003) have generally come to similar
conclusions. The E+A galaxies (Dressler \& Gunn 1992) or post-starburst 
galaxies (Couch \& Sharples 1987) are a puzzling exception. The
large numbers found by the MORPHs group (Dressler et al. 1997)
suggest that there was strong star formation activity in the recent
past in many galaxies (but see Balogh et al. 1999). A possible
explanation is that these galaxies have only recently arrived in the
cluster from much lower-density environments.
Indeed, field studies at low redshift have shown this type of
object to be more common in low-density regions than in clusters
(Zabludoff et al.\ 1996; Goto et al.\ 2003; Quintero et al. 2003). Therefore,
the greater numbers of E+A galaxies
found in high-redshift clusters may result from
the greater star formation activity of galaxies outside clusters
--- this idea gains strong support from
Tran et al.'s (2003) observations presented at the Symposium.

The next step is to compare the star formation rates in clusters cores at
different redshifts. Work is only just starting on this using emission-line
strengths (e.g., Ellingson et al. 2001), since it is essential to
control systematic uncertainties,
such as the aperture through which the star formation rate is measured.
However, extensive comparisons have been made on the basis of colors,
starting with Butcher \& Oemler (1978, 1984) and Couch \& Newell (1984).
These papers showed a startling increase in the numbers of blue galaxies
in $z>0.2$ galaxy clusters compared to the present day.
These results have been confirmed by more recent studies
(e.g., Rakos \& Schombert 1995; Margoniner et al.\ 2001), although the
effect of the magnitude limit and cluster selection play at least as
important a role as the redshift (Fairley et al.\ 2002).

There are two issues that complicate the comparison of the galaxies
in cluster cores, however. Firstly, we must be
careful how we select galaxies that are to be compared. Most
of the blue galaxies lie close to the photometric completeness
limit. These galaxies will fade by up to 1~mag if star formation is
turned off, and thus they are not directly comparable to the
red-galaxy population selected at the same magnitude limit (Smail et
al. 1998; Kodama \& Bower 2001). Secondly, we are observing galaxy
clusters in projection. There is little doubt that the field galaxy
population at intermediate redshift is much bluer than in the local
Universe (Lilly et al.\ 1995; Madau, Pozzetti, \& Dickinson 1998);
 thus, although a
small level of contamination by field galaxies has little influence on
the overall color distribution, the same contamination will have a
much bigger impact on the distribution at intermediate redshift. This
problem is only partially eliminated if a complete sample of
galaxy redshifts is available since the velocity dispersion of the
cluster makes it impossible to distinguish cluster members from
``near-field'' galaxies that are close enough to the cluster to be
indistinguishable in redshift space (Allington-Smith et al.\ 1993;
Balogh et al.\ 1999; Ellingson et al.\ 2001).  This idea is reinforced
by experiments with numerical simulations. Galaxies can be associated
with dark matter particles, and then ``observed'' to measure the extent
to which radial information is lost. Diaferio et al.\ (2001) found that
a contamination of 10\% can easily occur; furthermore, since most of
the contaminating galaxies are blue (and in these models most
genuine cluster galaxies are red), the fraction of blue galaxies
can then be boosted by 50\%. Despite this,
Ellingson et al.\ (2001) conclude that the rate at which clusters are being
built up must also be higher in the past in order for this explanation
to work. Kauffmann (1995) shows that there is good theoretical justification
for this.

It will be interesting to see if the evolution in the colors
of the cluster population are consistent with the evolution in the
emission-line strengths. We might expect to see a difference
because of the different time scales probed by colors and by emission lines.
For example, if galaxies that fall into the cluster have their star formation
quickly suppressed, they will remain blue (in the Butcher-Oemler sense) for a
significant period after the line emission subsides (Ellingson et al.\
2001).  Combining these factors, it seems quite possible to accommodate
both weak evolution in emission-line strength and more rapid evolution in
the colors of cluster galaxies.

\section{The Other Axis: Density}

\subsection{The Cluster Outskirts}

So far we have been discussing the properties of galaxies within the
cores of clusters, but the dependence on density (or, nearly
equivalently, cluster-centric radius) provides
another axis over which to study galaxy properties. We have seen that
star formation is strongly suppressed in the cores of rich clusters ---
but at what radius do the galaxies become more like the field? We
should also realize that it might be better to compare galaxy
properties with their local densities (Dressler 1980; Kodama et
al. 2001), as the large-scale structure surrounding clusters may have
the dominant impact on galaxy evolution.

\begin{figure*}[t]
\centering
\includegraphics[width=0.49\columnwidth,angle=0,clip]{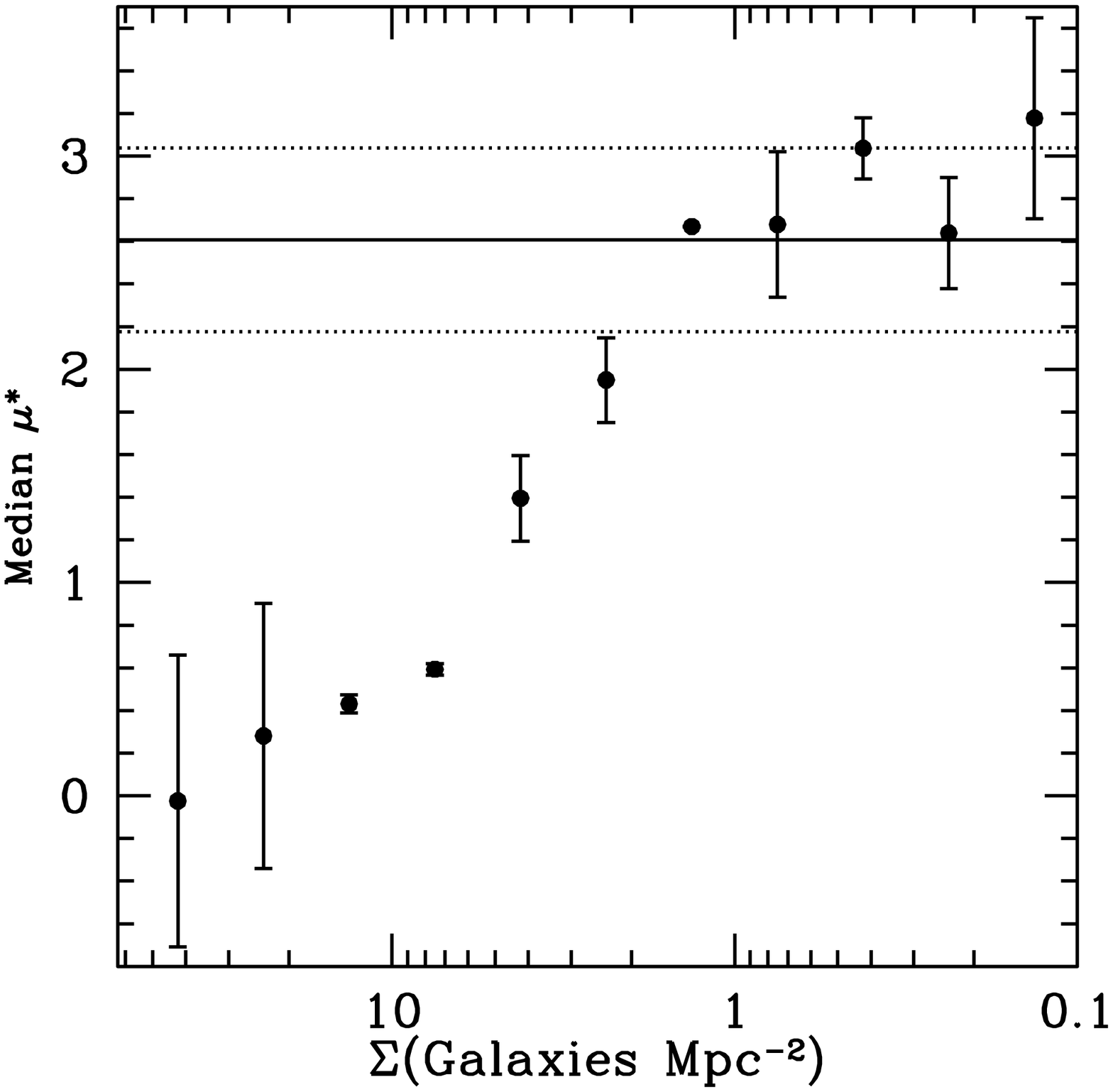}
\includegraphics[width=0.49\columnwidth,angle=0,clip]{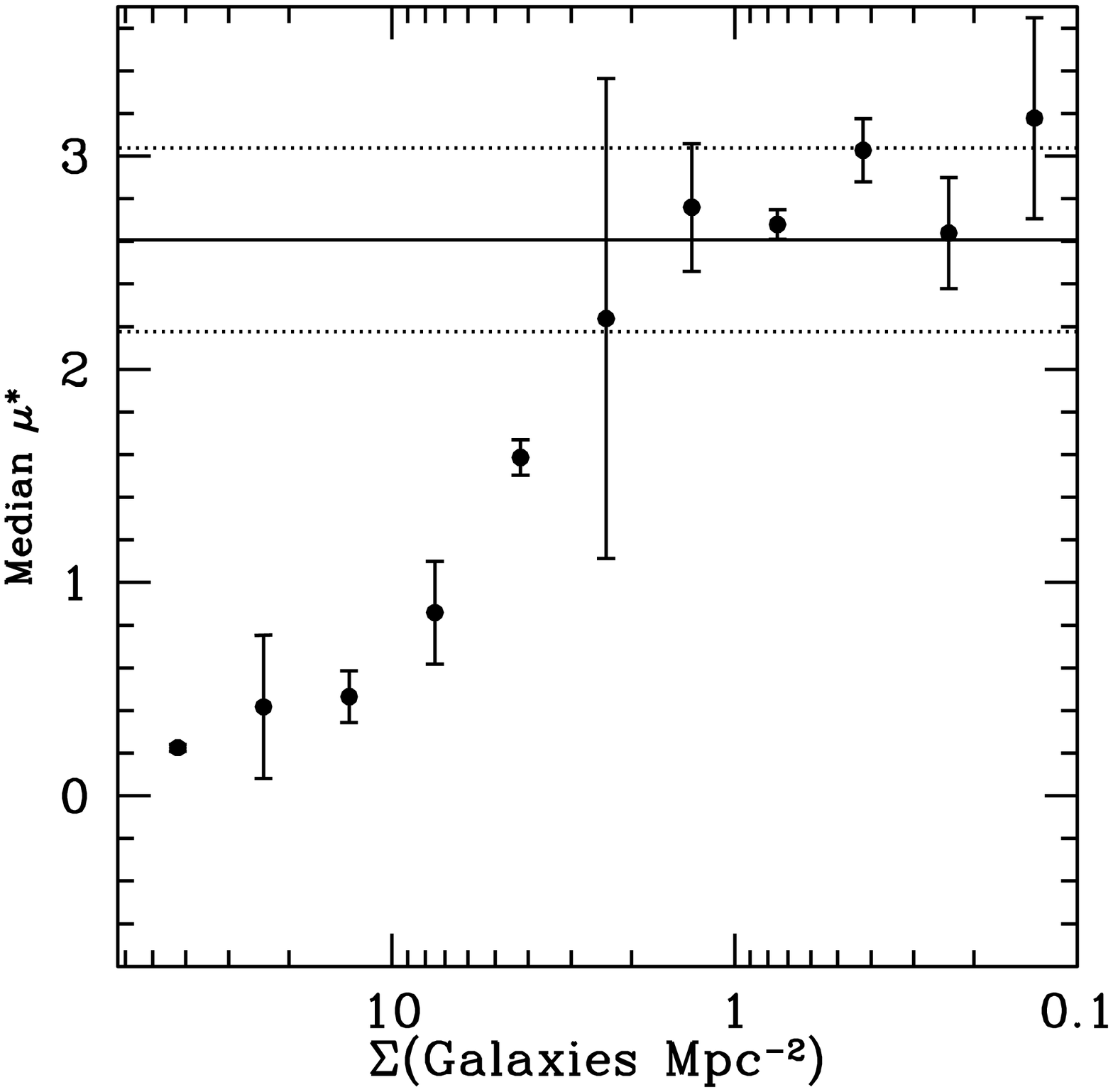}
\vskip 0pt \caption{
The median star formation rate as a function of local density from galaxies 
around clusters in the 2dF survey (based on Lewis et al. 2002). The {\it left}
panel shows the star formation rate of all galaxies in the sample; the 
{\it right} panel shows the effect of removing galaxies within 2 virial radii 
of the cluster center. Error bars are jackknife estimates.  The relations
are amazingly similar, showing that the local density is more important
than the overall mass of the group or cluster.}
\label{fig:sfr_density}
\end{figure*}

One of the first steps at studying galaxies in the transition zone
around clusters were made with the CNOC2 survey
(Balogh et al. 1999). They showed that there was a strong radial dependence
in the star formation rate, but that the star formation rate had not yet
reached the field value even at $r\approx  r_{\rm vir}$.
The Sloan Digital Sky Survey (SDSS) and 2dF galaxy redshift survey surveys 
have allowed us to make a huge leap
forward in this respect. In the local Universe, we are able to map
galaxy star formation rates, using the complete redshift information
to eliminate contamination by interlopers. In this section we will
concentrate on what we have learned from the 2dF survey (Lewis et al.
2002), but the results from the SDSS give very consistent answers
(G\'omez et al. 2003).
Figure~\ref{fig:sfr_density} shows the median star formation
rate as a function of local density. What is remarkable in this plot
is that there is quite a sharp transition between galaxies with
field-like star formation rates at $\Sigma < 1 \Mpc^{-2}$ and galaxies with low
star formation rates comparable to cluster cores ($\Sigma > 7 \Mpc^{-2}$). The
switch is complete over a range of less than 7 in density.

The density at which the transition occurs corresponds to the
density at the virial radius. If star formation is plotted against radius, the
transition is considerably smeared out, but does occur at around the
cluster virial radius --- well outside the core region on which a lot
of previous work has been focused. The 2dF galaxy redshift survey sample is sufficiently large that we can remove the cluster
completely from this diagram. By only plotting galaxies more than 2 
virial radii from the cluster centers, we concentrate on the filaments
of infalling material. The correlation with local density is shown in
Figure~\ref{fig:sfr_density}. Amazingly, the relation hardly changes
compared to the complete cluster diagram.

This is a great success: we have identified the region where galaxy
transformation occurs! It is in the infalling filaments (consisting
of chains of groups) where galaxies seem to change from star-forming,
field-like galaxies to passive, cluster-like objects. Of course, it is
tempting to associate the transformation in star formation rate with a
transformation from late- to early-type morphology.
Unfortunately, this test cannot be undertaken with
the available 2dF data, but we can expect clearer results from SDSS.

\begin{figure*}[t]
\centering
\includegraphics[width=0.75\columnwidth,angle=0,clip]{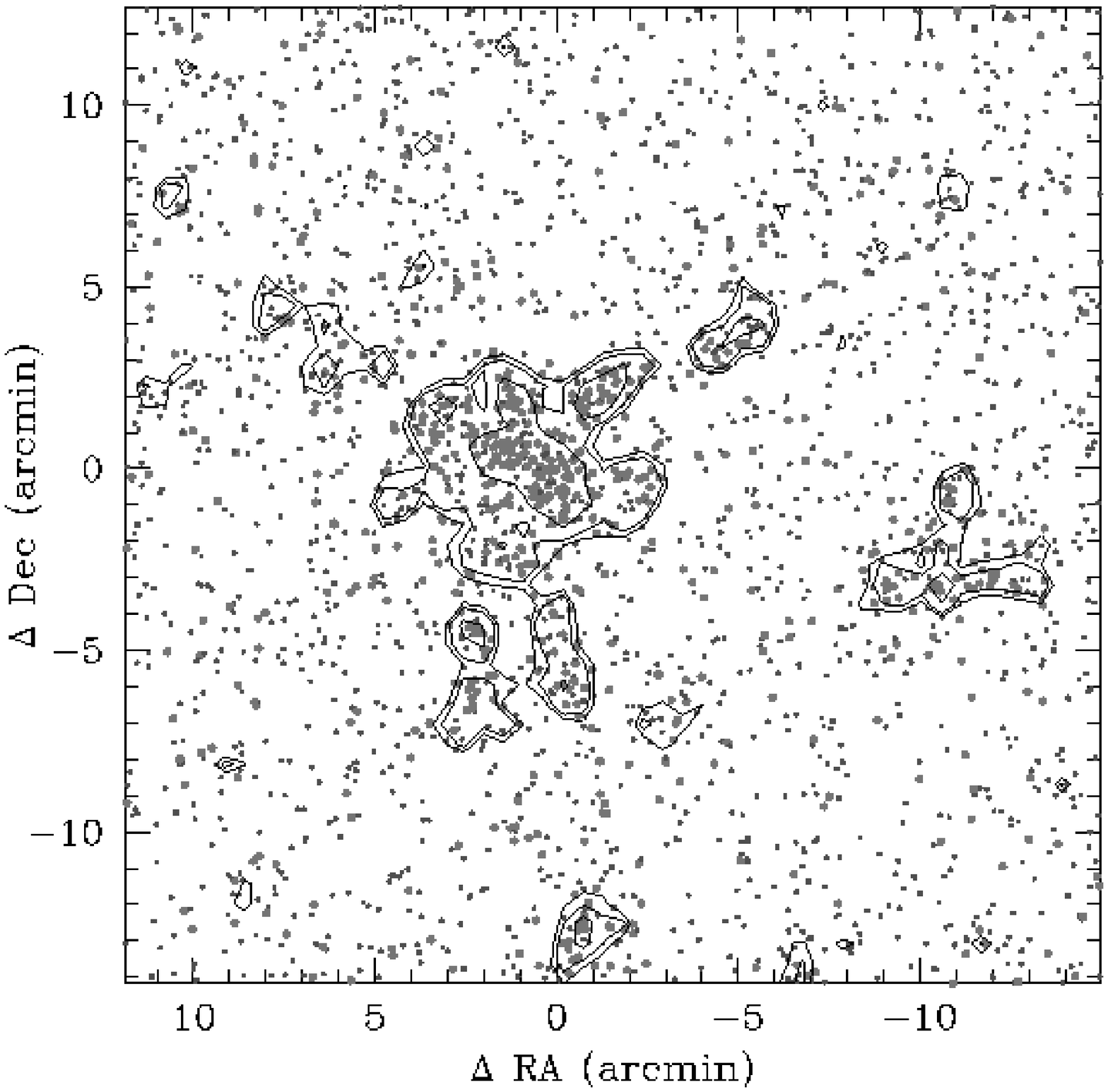}
\vspace*{-0.5cm}
\includegraphics[width=0.75\columnwidth,angle=0,clip]{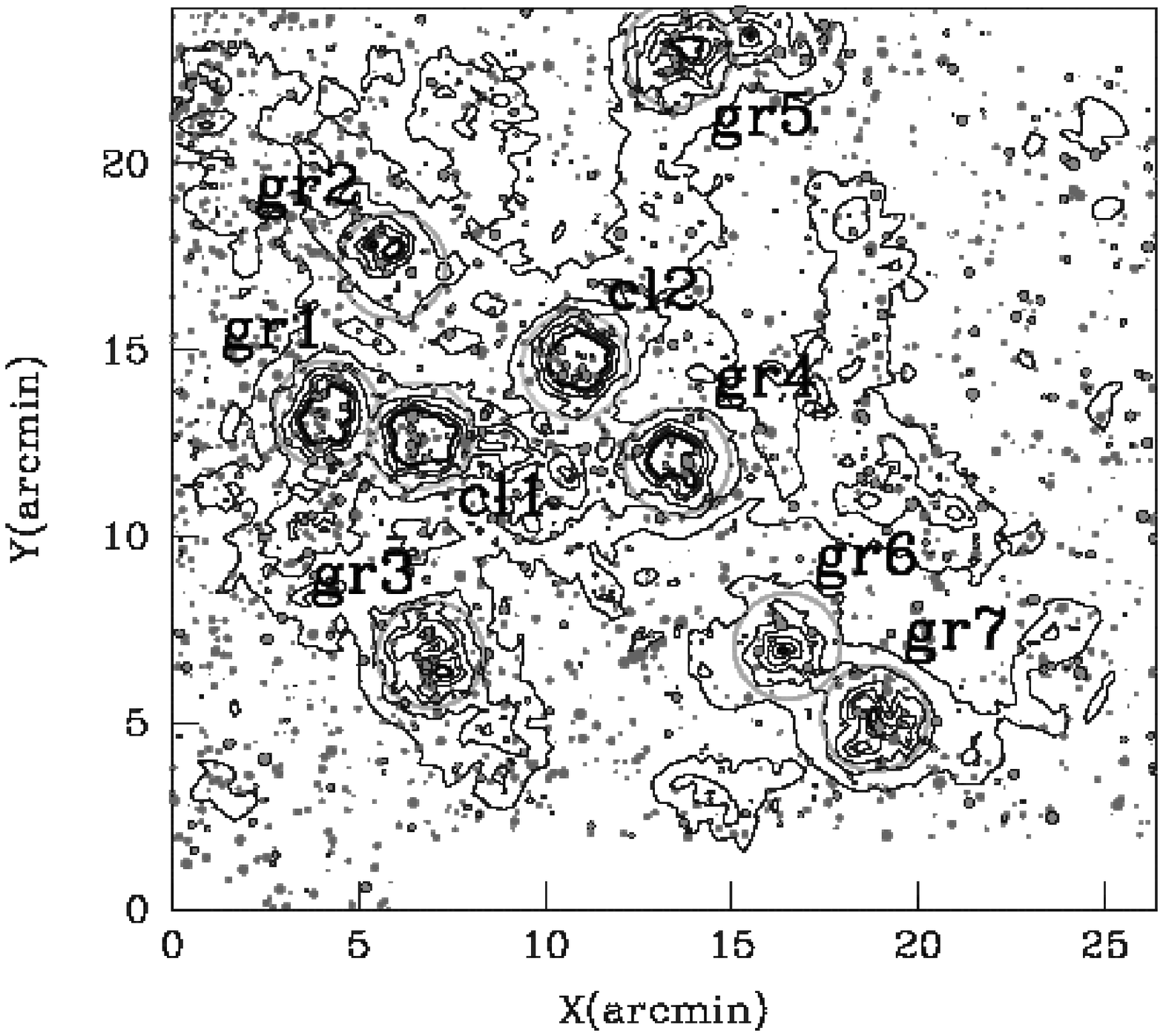}
\vspace*{+0.5cm}
\vskip 0pt \caption{
{\it Top}:  Contour lines pinpoint the
transition zone around the rich cluster A851
at $z=0.41$  studied by Kodama et al.\ 2001.   {\it Bottom}:
$z=1.27$ groups tracing the large-scale structure surrounding the
Lynx clusters (based on Nakata et al.\ 2003). These figures illustrate
how photometric methods can be used to reduce contamination rates
in the outskirts of distant clusters so that the galaxy population
can be studied there.}
\label{fig:a851_lynx}
\end{figure*}

What happens at higher redshift? In fact, the first claim of a
sharp transition in galaxy properties was made by Kodama et al. (2001) for
the distant cluster A851 at $z=0.41$ (top panel in
Fig.~\ref{fig:a851_lynx}). Kodama et al. (2001) used photometric
redshifts to eliminate foreground objects, and thus to reduce
contamination of the cluster members to a level that allowed the
color distribution to be studied in the outer parts of the cluster.
Their results show an amazing transition in color.
Direct comparison with the local clusters is difficult, however, as
the magnitude limits are very different (Kodama et al.'s photometric data
reach much fainter than the local spectroscopic samples), but
G\'omez et al.\ (2003) concluded that the threshold seen by Kodama et al. (2001)
 was at a
significantly higher local density. Perhaps dwarf galaxies are more
robust to this environmental transformation; we are not going to
attempt to cover this issue.

A number of researchers are now engaged in spectroscopic programs to study the
transformation threshold in higher-redshift systems. The results of
Treu et al.\ (2003) are perhaps the most advanced. They also have the
advantage of panoramic WFPC2 imaging that will allow them to compare
the transformation of galaxy morphology (see Treu 2003).

The highest redshifts that can be studied require a combination of
photometric preselection of objects for spectroscopy. Nakata et
al.\ (2003) have used the photometric technique to map the large-scale
structure around the Lynx cluster at $z=1.27$ (lower panel in
Fig.~\ref{fig:a851_lynx}),
and similar techniques are described by Demarco et al.\ (2003).
These groups identify several candidate filaments; spectroscopy of these
regions is now underway.

\subsection{Galaxy Groups}

Returning to the local Universe, it is interesting to see if we can
probe the properties of galaxies in groups directly. A lot of work has
been carried out looking at small samples of groups selected from the
CfA redshift survey (Geller \& Huchra 1983;  Moore, Frenk, \& White 1993),
from the Hickson compact group catalog (Hickson, Kindl, \& Auman 1989), and
also from X-ray surveys (Henry et al.\ 1995; Mulchaey et al. 2003).

In the era of the 2dF and SDSS redshift surveys, we can construct
robust catalogs containing thousands of groups (Eke et al. 2003).
It is interesting to
compare the star formation rate in the groups as a function of local
density, with the relation found in clusters.  The relation

\clearpage

\begin{figure*}[t]
\centering
\includegraphics[width=0.49\columnwidth,angle=0,clip]{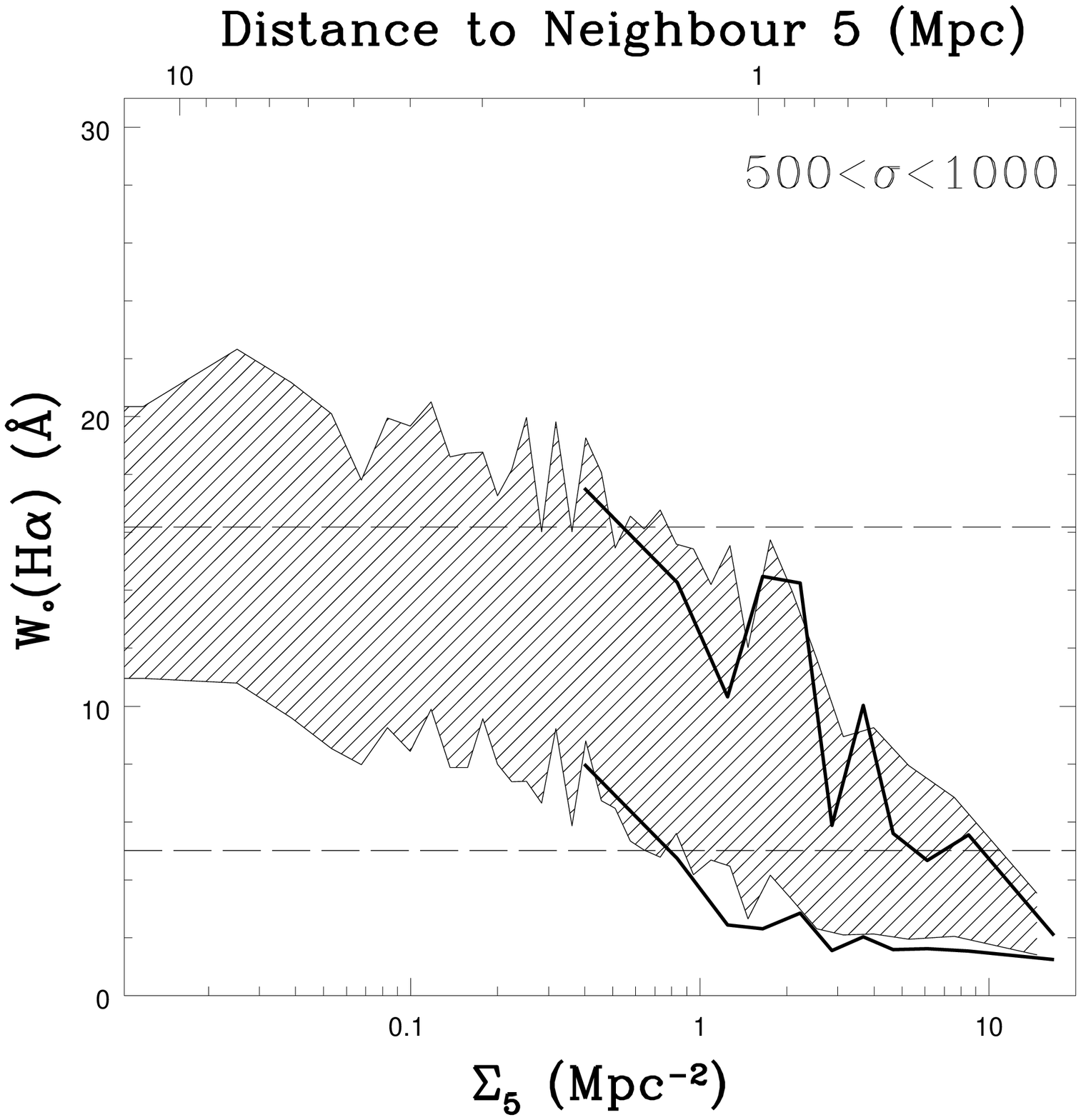}
\includegraphics[width=0.49\columnwidth,angle=0,clip]{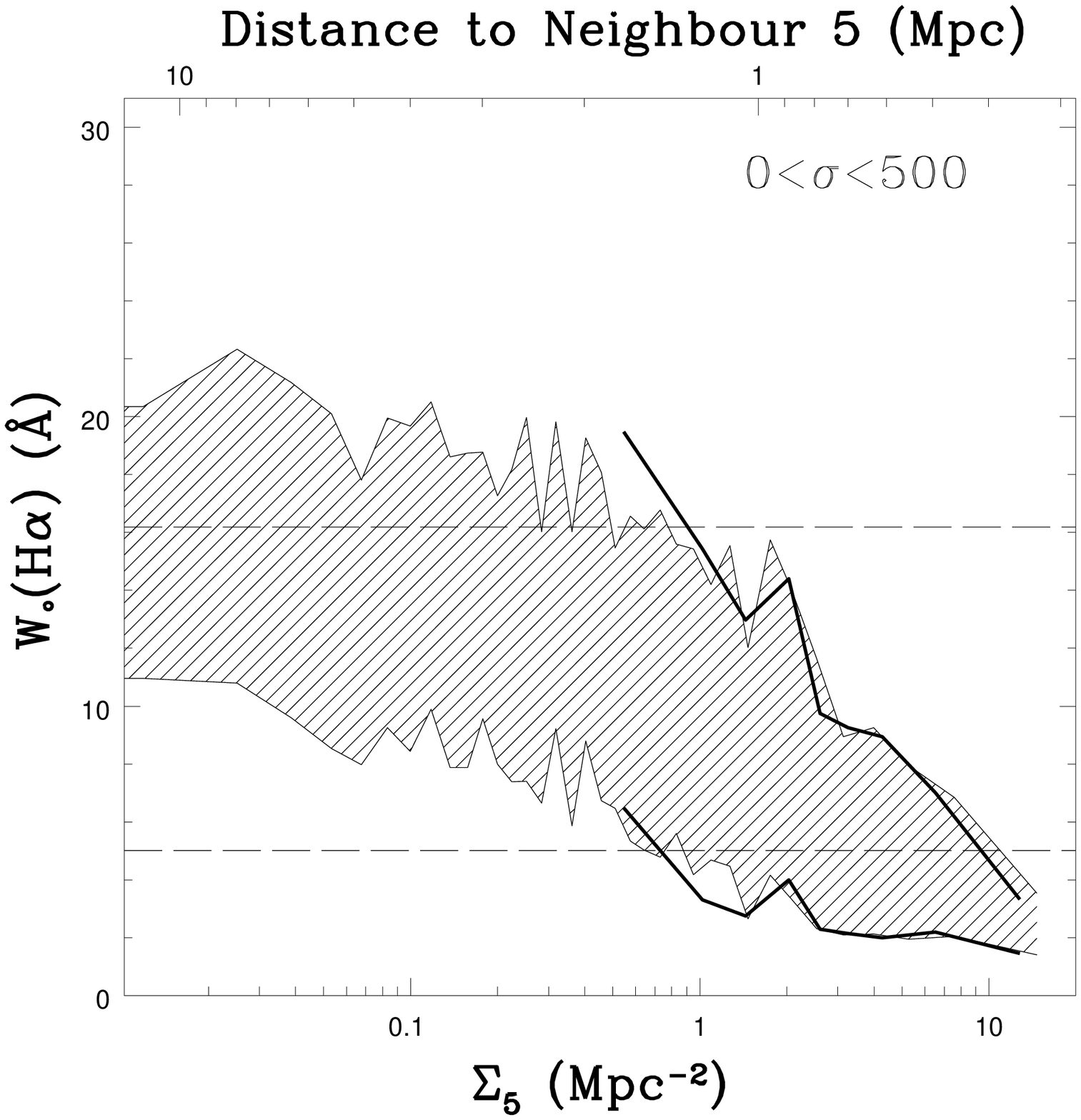}
\vskip 0pt \caption{
The star formation rate as a function of density, comparing
groups of galaxies with clusters. The upper and lower horizontal dashed lines  
show the 75$^{\rm th}$ percentile and the median of the equivalent widths.  
The hashed region shows the relation for the complete sample, while the solid 
line shows the relation for systems with $500\,\kms\ <\sigma<1000 \,\kms$ 
({\it left}) and $\sigma <500 \,\kms$ ({\it right}). The dependence on local 
density is identical irrespective of the velocity dispersion of the whole 
system. Figure based on Balogh et al.  (2003).} 
\label{fig:group_sfr}
\end{figure*}

\clearpage

\noindent
for the
2dF survey is shown in Figure~\ref{fig:group_sfr} (Balogh et al. 2003).
The panels show the effect of selecting systems on the basis of their velocity
dispersion. There is actually very little difference between the trends. The
galaxies in dense regions suffer the same suppression of their star formation
rate, regardless of the system's total mass. It is also possible to show
that the groups in the infall regions of clusters show the same
pattern as isolated groups.  We have to conclude that the suppression
of star formation is very much a local process. This is an important clue
to distinguish between the different transformation mechanisms.

\begin{figure*}[t]
\includegraphics[width=1.00\columnwidth,angle=0,clip]{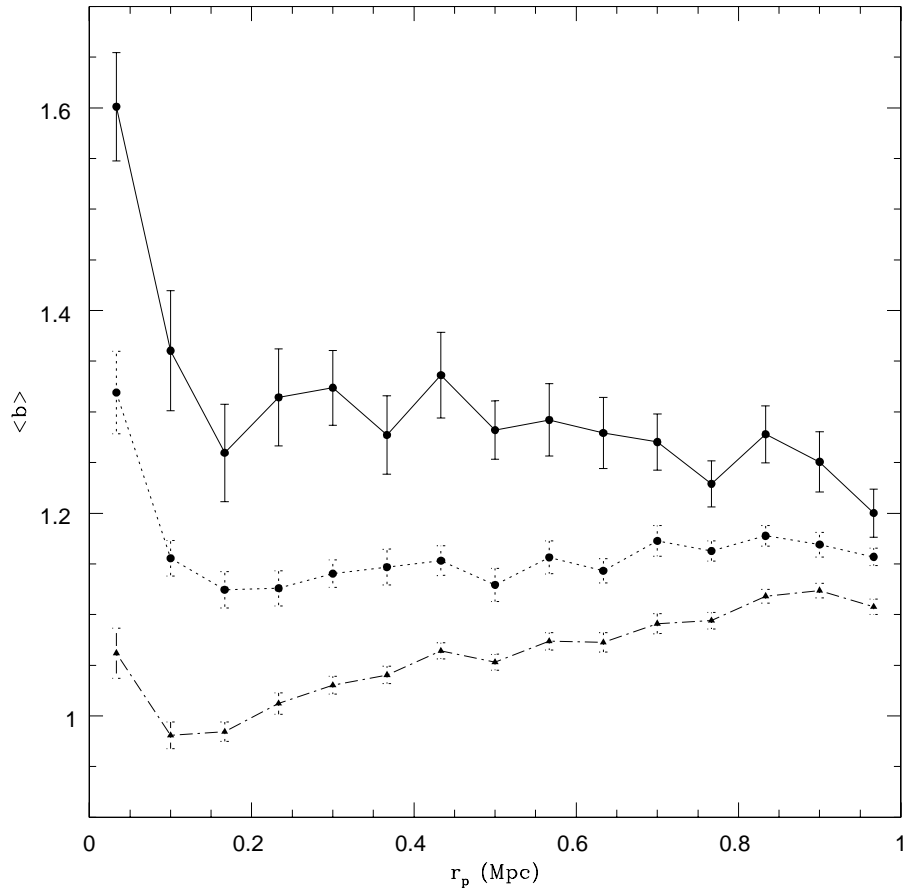}
\vskip 0pt \caption{
The star formation rate of galaxy pairs ($b$) as a function of
separation. The different line styles distinguish the spectral type
of the first galaxy of the pair, with the sequence of dot-dash, dotted, and
solid lines showing the effect of restricting the sample to more
and more active central objects. A strong enhancement is only seen
when the separation is less than 100~kpc. The figure is based on
Lambas et al.\ (2003).}
\label{fig:sfr_sepn_vdisp}
\end{figure*}

Interestingly, in the local Universe, there is little evidence for the
environment producing a rise in the star formation rate above the field
value. The only exception to this appears to be the close, low-velocity
encounters of isolated galaxies (Barton, Geller, \& Kenyon 2000; 
Lambas et al.\ 2003).
Figure~\ref{fig:sfr_sepn_vdisp} shows the star formation rate
as a function of separation for systems of different total velocity
dispersion. A spike in the median star formation rate appears only in
the smallest bin of the first panel.  It will be interesting to study
this trend within groups and clusters (Balogh et al. 2003).

\begin{figure*}[t]
\includegraphics[width=1.00\columnwidth,angle=0,clip]{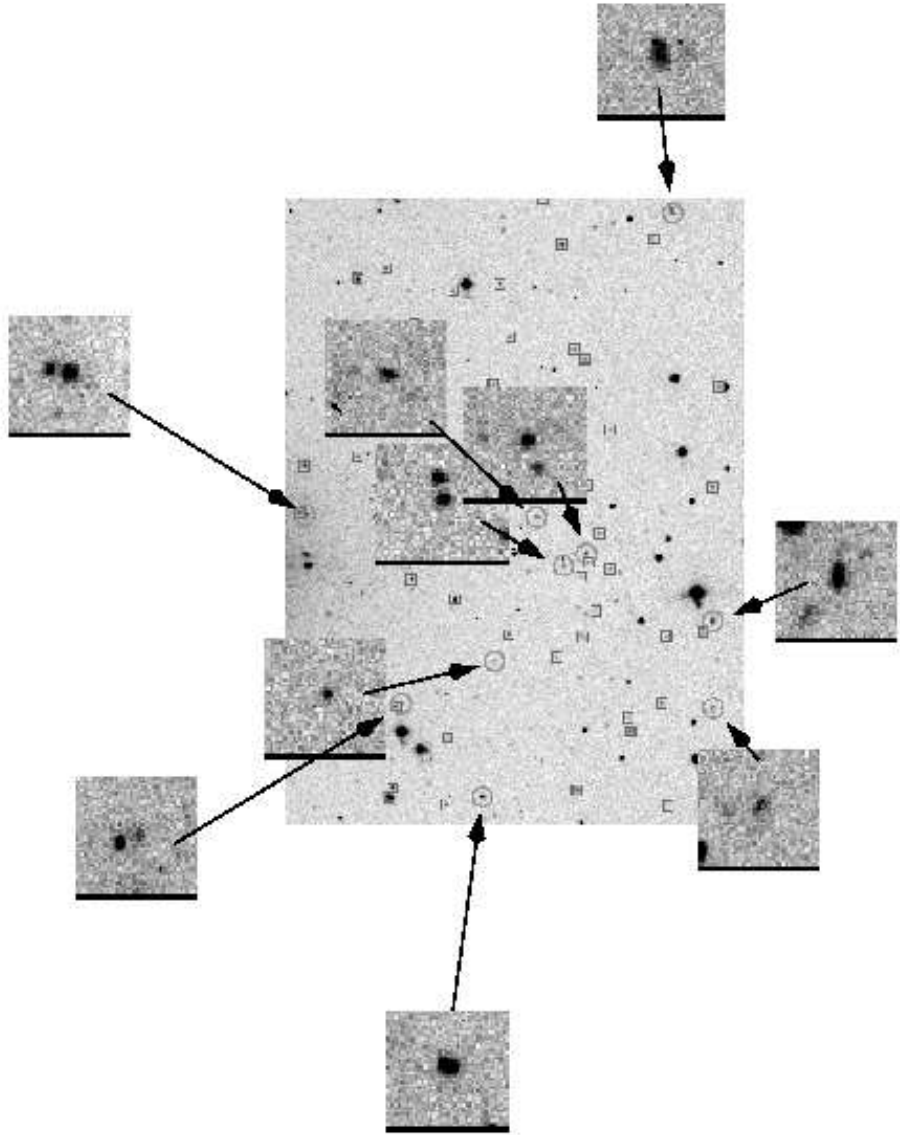}
\vskip 0pt \caption{
A sample group from the CNOC2/Magellan group survey. This
poor group, containing 10 members, is at $z=0.393$.  }
\label{fig:g134_overlay}
\end{figure*}

One of the next goals is to extend studies of groups to
higher redshifts.  The first steps in this direction were made by
Allington-Smith et al.\ (1993). They used radio galaxies to pick out
galaxy groups at redshifts up to 0.5. By stacking photometric catalogs,
they showed that the galaxy populations of rich groups 
($N_{0.5}^{-19}>30$)\footnote{Group richness defined as the number of galaxies 
with $M_V \le -19$ mag within a 0.5 Mpc radius of the radio galaxy 
($H_0$ = 50 km s$^{-1}$ Mpc$^{-1}$ and $q_0$ = 0 assumed).}
became increasingly blue with redshift, while poorer groups contained
similar populations of blue galaxies at all redshifts.
A survey of redshift-space selected groups at intermediate redshift became
possible with the CNOC2 redshift survey. Carlberg et al.\ (2001) report a
statistical sample of 160 groups out to redshift 0.4. On the Magellan
telescopes, we have been following up the systems at $z>0.3$ in order to
determine the complete membership and measure total star formation rates.
Figure~\ref{fig:g134_overlay} shows the membership of a sample group.
The initial results are exciting --- star formation in many galaxies
are more comparable
to the surrounding field values.  If these results are confirmed as we
derive more redshifts and improve the group completeness, it represents a
very interesting change from the properties seen in the 2dF groups.
At higher redshifts, a tantalizing glimpse of the properties of a few
groups can be obtained from the Caltech redshift survey (Cohen et al.\
2000).

\section{What Does It All Mean?}

\subsection{The Mechanisms Driving Galaxy Evolution}

\begin{figure*}[t]
\includegraphics[width=1.15\columnwidth,angle=0,clip]{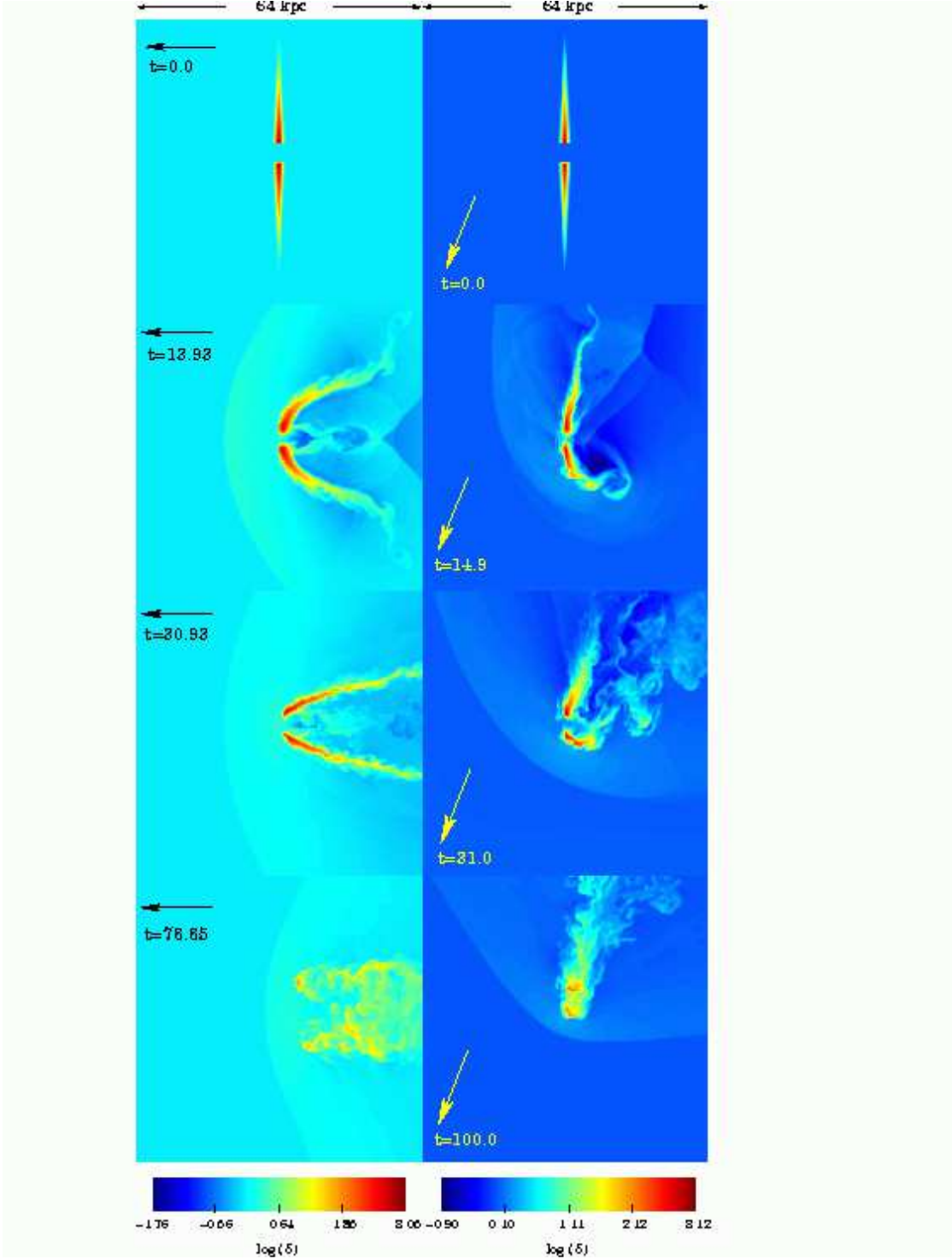}
\vskip 0pt \caption{
A numerical simulation of the ram pressure stripping of a spiral galaxy, based
on Quilis et al. (2000).  The left and right panels compare the effect when
the galaxy is face-on and almost edge-on.}
\label{fig:ram_pres}
\end{figure*}

The mechanisms that have been proposed to drive galaxy evolution in
dense environments can be broadly separated into three categories.
\begin{itemize}
\item 
{\it Ram pressure stripping}. Galaxies traveling through a dense
intracluster medium suffer a strong ram pressure effect that sweeps
cold gas out of the stellar disk (Gunn \& Gott 1972; Abadi, Moore, \& Bower 
1999; Quilis, Moore, \& Bower 2000).  The issue with this mechanism is whether 
it can be effective outside dense, rich cluster cores where the galaxy 
velocities are very high and the intracluster medium is very dense. Quilis et 
al.\ (2000) found that
incorporating holes in the galaxy H~{\sc i} distribution made galaxies
easier to strip (Fig.~\ref{fig:ram_pres}), but it still required clusters
more massive than the Virgo cluster to have a great effect.

\item 
{\it Collisions and harassment}. Collisions or close encounters between
galaxies can have a strong effect on their star formation rates. The
tidal forces generated tend to funnel gas toward the galaxy center
(Barnes \& Hernquist 1991; Barnes 2002; Mihos 2003). It is likely that this will fuel
a starburst, ejecting a large fraction of the material (Martin 1999).
Gas in the outer parts of the disk, on the other hand, will
be drawn out of the galaxy by the encounter. Although individual
collisions are expected to be most effective in groups because the
velocity of the encounter is similar to the orbital time scale within
the galaxy, Moore et al.\ (1996) showed that the cumulative effect of
many weak encounters can also be important in clusters of galaxies.

\item 
{\it Strangulation}. Current theories of galaxy formation suggest that
isolated galaxies continuously draw a supply of fresh gas from a hot,
diffuse reservoir in their halo (Larson, Tinsley, \& Caldwell
1980; Cole et al. 2000). Although the reservoir is too cool and diffuse to be easily
detected (Benson et al.\ 2000; Fang, Sembach, \& Canizares 2003), 
this idea is supported by
the observation that 90\% of the baryonic content of clusters is in
the from of a hot, diffuse intracluster medium. The baryon reservoir in galaxy halos
is entirely analogous. When an isolated galaxy becomes part of a
group, it may loose its preferential location at the center of the
halo and thus be unable to draw further on the baryon reservoir. Without
a mechanism for resupplying the material that is consumed in
star formation and feedback, the galaxies' star formation rate will
decline. The exact rate depends on the star formation law that is used
(Schmidt 1959; Kennicutt 1989) and on whether feedback is strong
enough to drive an outflow from the disk.

\end{itemize}

Semi-analytic models (e.g., Cole et al. 2000) generally
 incorporate only the third of these mechanisms.
The observational data strongly suggest that the ram pressure stripping
scenario cannot be important for the majority of galaxies. As we have
seen, the suppression of star formation seems to occur well outside of
the clusters and is equally effective in low-velocity groups, which do not 
possess a sufficiently dense intracluster medium.  Distinguishing
between the remaining two scenarios is rather harder, since they have
similar dependence on environment. Indeed, they may
both play a role. The key difference is the time scale on which they operate:
collisions are expected to produce changes in galaxy
properties on short time scales ($\sim 100 \,\hbox{Myr}$), while
the changes due to strangulation are much more gradual ($>1\,\hbox{Gyr}$).
The time scale for harassment is less well defined; while the individual
encounters may induce short-lived bursts of star formation, the overall
effect may accumulate over several Gyr. The radial gradients that we
observe appear to prefer long time scales and, hence, a mechanism like
strangulation or harassment (Balogh, Navarro, \& Morris 2000).  To make further
progress in this area, we need to compile detailed observations of
galaxies that are caught in the transition phase.  In particular,
morphological measurements will provide another important distinction
(e.g., McIntosh et al.\ 2003).

\clearpage

\clearpage

\begin{figure*}[t]
\includegraphics[width=0.49\columnwidth,angle=0,clip]{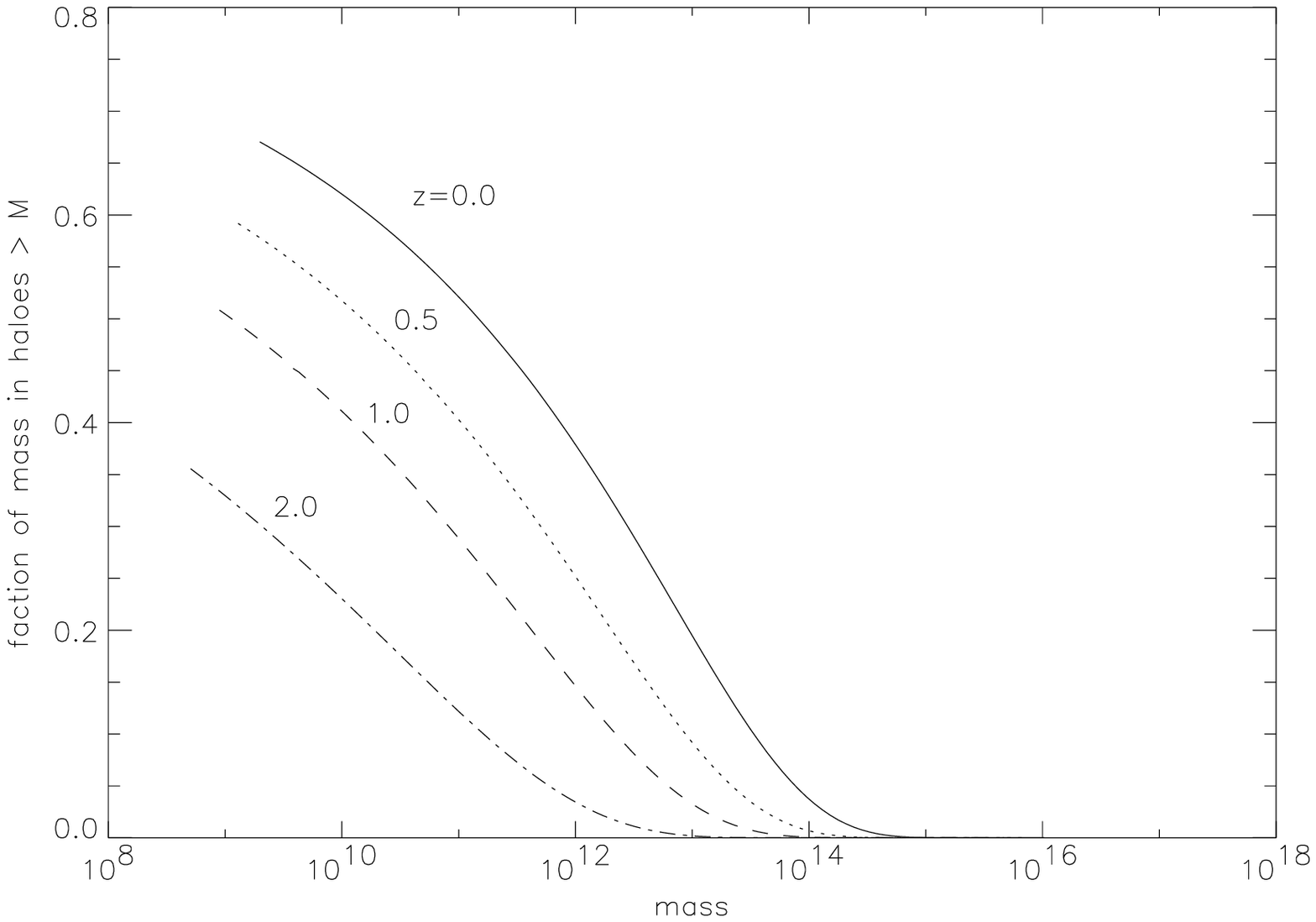}
\includegraphics[width=0.49\columnwidth,angle=0,clip]{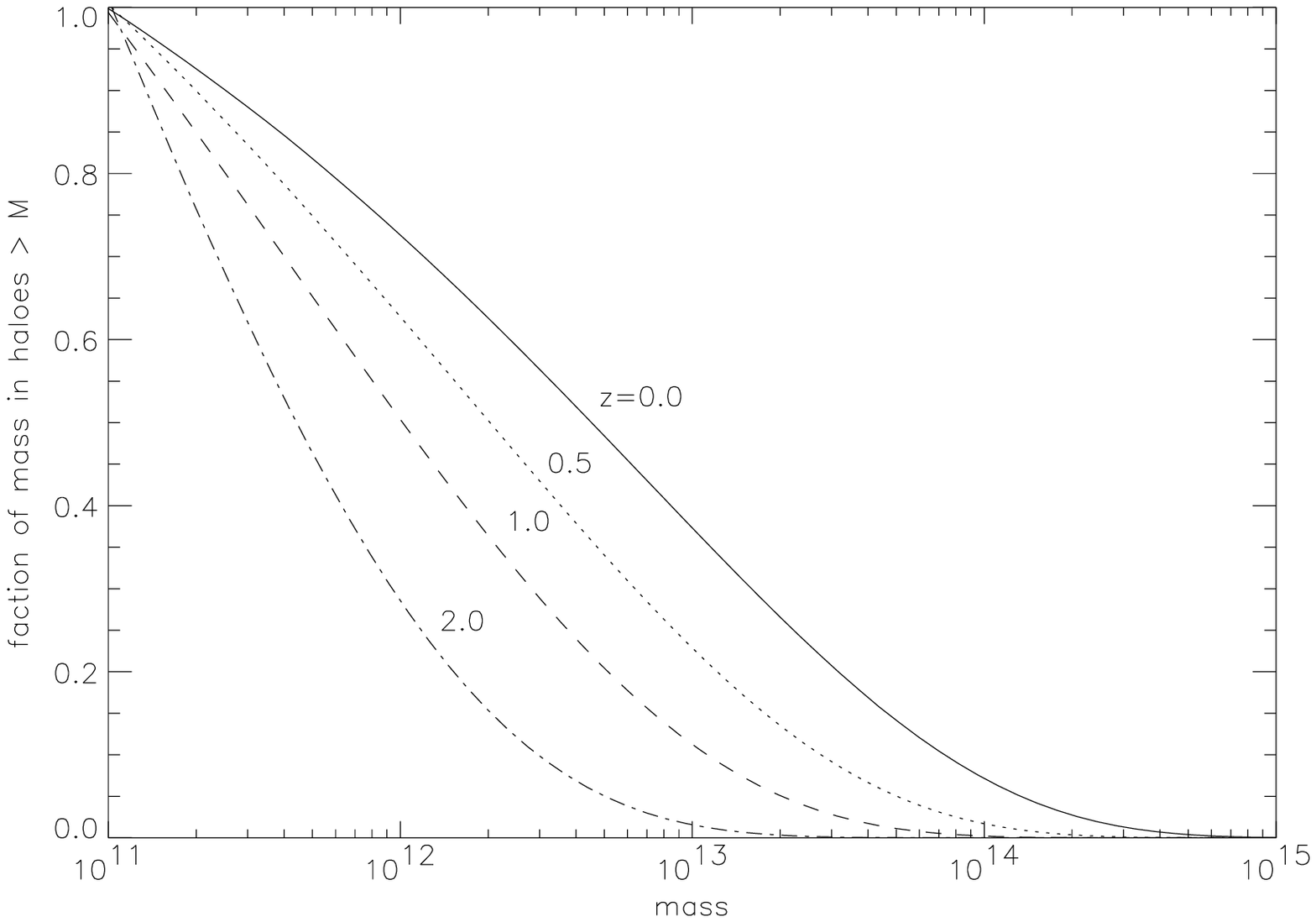}
\vskip 0pt \caption{
The fraction of mass in the Universe bound
into objects more massive than the mass scale on the horizontal axis. The
{\it left} panel shows the curves with their overall normalization. Note
that most of the mass is actually in subgalactic halos even in the
$z=0$ Universe.  The {\it right} panel shows the curves renormalized at
$10^{11} \Msol$. Between $z=1$ and the present day, there is a dramatic
increase in the fraction of the renormalized mass contained in groups.
The issue is to determine whether the growth of structure is responsible
for the decline of the cosmic star formation rate.}
\label{fig:mass_frac}
\end{figure*}

\subsection{The Star Formation History of the Universe}

In addition to its own intrinsic interest, one reason for studying the
impact of the environment on galaxy evolution is to understand the
down-turn in the cosmic star formation rate. Studies of the global
star formation rate show a decline of a factor 3--10 since the peak of
star formation at $z=1$--2 (Lilly et al.\ 1995;
Madau et al.\ 1998; Glazebrook et al.\ 1999; Wilson et al.\ 2002). We can
simplify the possible explanations into two alternative hypotheses: (1) 
the down-turn is caused by galaxies running out of a supply of
material for star formation, or (2) it is driven by the growth of the
mass structure of the Universe.  
In the first scenario, the down-turn
is intrinsic to the galaxy population; in the second, it is caused by
the changing environments of galaxies. Of course, the truth probably
lies somewhere in between. In popular ``semi-analytic'' models (for
example, Kauffmann et al.\ 1999; Somerville \& Primack 1999;
Cole et al. 2000),
the decline occurs because galaxies that are not at the centers of
their halo potential cannot accrete fresh gas, because this supply is
being switched off by the growth of the halo mass.

It is useful to consider a toy model to investigate whether the growth of
structure in scenario (2) is sufficiently rapid to be viable. Defining
groups and clusters by their mass (as discussed in the Introduction),
we can plot the fraction of the mass in ``groups'' and ``clusters'' as
a function of redshift. Figure~\ref{fig:mass_frac} shows
that this fraction is always small, but that the fraction of mass in groups
changes by a factor of 3--5 since the peak of cosmic star formation history.
Of course, this model is highly simplified, so one should only treat it
as an illustration of the idea.

\begin{figure*}[t]
\includegraphics[width=0.7\columnwidth,angle=270,clip]{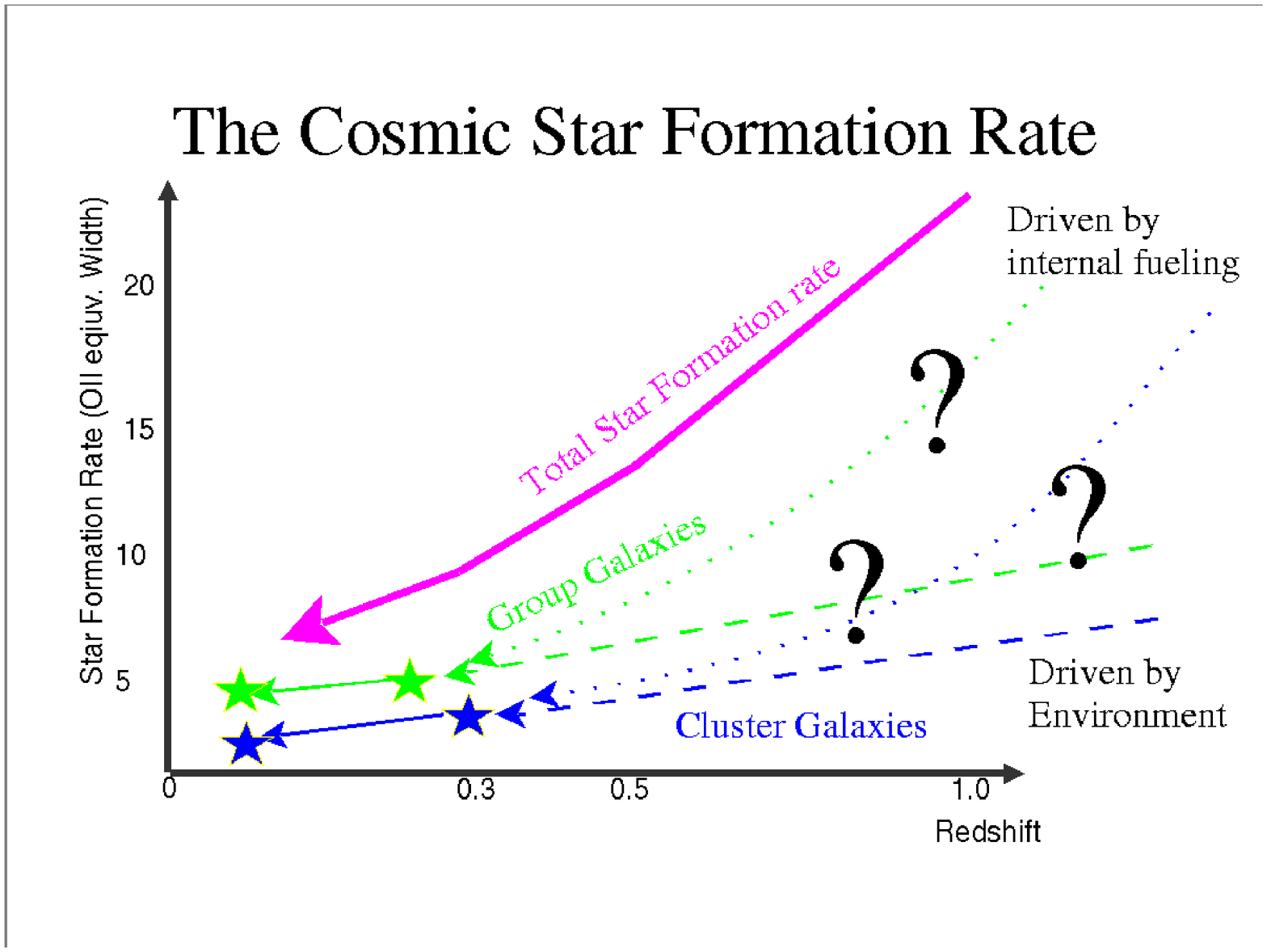}
\vskip 0pt \caption{
   An illustration of how we might expect the cosmic
  star formation history to vary between galaxies in different
  environments. As we look back in time, do galaxies in clusters and groups
  mirror the trend in the total star formation rate, or is star formation in
  galaxies in dense environments suppressed to similarly low levels at all
  redshifts?  The plot hints at how we might interpret these different
  environment-dependent histories. Note that the curves drawn in this
  figure refer to the star formation rates of galaxies observed to be
  in groups and clusters at the redshift shown. The history of an individual
  galaxy (which cannot be observed, but may be extracted from simulations)
  will jump from one track to another as it is accreted from the field
  into groups and filaments and finally into the cluster core.}
\label{fig:env_madau_plot}
\end{figure*}

By observing groups and clusters of galaxies at different redshifts,
we can hope to combine the data from different environments to make an
``environmental Madau plot'' where we break the contribution to the
star formation rate down into its contributions from different
environments.  This is a task that is becoming more easily within our
grasp. Some suggestions for how the plot might look are shown in
Figure~\ref{fig:env_madau_plot}.
An extension to the concept is to treat galaxy formation as an inverse
problem. We have good models for how the dark matter halos of
galaxies evolve and combine, so we can connect together galaxies in
groups at $z=1$ with galaxies in clusters at $z=0$. By combining our
observations of galaxies in different environments at different
redshifts with these numerical models, we can solve for the star
formation histories of galaxies along this trajectory.

\subsection{A Closing Thought}

Throughout this review we have taken it for granted that the reason
why galaxies in clusters end up looking different from galaxies in the
field is causally related to their present environment --- that
there is a definite moment of transformation when field galaxies are
transformed into passive cluster/group-like objects. But it is worth
pausing to consider whether this is necessarily true. We wonder if it
is still possible to believe that the cluster galaxies initially form
differently (e.g., in a much more rapid collapse) and then just
happen to end up in clusters (i.e., the ``nature'' scenario). In this case there is no causal
connection.  For example, if we ask the question ``where are the
Lyman-break galaxies now?'', numerical simulations have shown that
these galaxies are now preferentially located in cluster cores
(Governato et al. 1998). This would seem to support the ``nature''
conjecture. However, we do not believe that the converse is necessarily true.
For example, if one selects all the galaxies in a present-day cluster and
ask in what environment (halo mass) they were located at $z=2$, one will find
that there is a huge spread in the distribution of environments, and that it is not straightforward to
distinguish this histogram from the corresponding histogram for
galaxies identified in the present-day field. Given that
the correspondence between the present-day environment and its
environment at $z=2$ is relatively weak, we find it difficult to see how
galaxies can ``predict'' their present-day environment at the
time they are being formed. Clearly, this argument needs to be placed
on a stronger footing by combining simulations and semi-analytic
techniques to accurately define the environment of galaxies at each
epoch (Diaferio et al.\ 2001; Springel et al.\ 2001; Helly et al.\ 2003).

\vspace{0.3cm}
{\bf Acknowledgments}.
RGB thanks the Leverhulme Foundation for its support, and MLB thanks PPARC for
its support. This work made extensive use of the NASA's Astrophysics Data
System online bibliographic database, and the ArXiv/Astro-ph electronic
preprint server.


\begin{thereferences}{}

\bibitem{AMB99}
Abadi, M.~G., Moore, B., \& Bower, R.~G. 1999, \mnras, 308, 947

\bibitem{} 
Allington-Smith, J.~A., Ellis, R.~S., Zirbel, E.~L., \& Oemler, A. 1993,
\apj, 404, 52

\bibitem{} 
Balogh, M. L., et al. 2003, in preparation

\bibitem{} 
Balogh, M.~L., Bower, R.~G., Smail, I., Ziegler, B.~L., Davies, R.~L.,
Gaztelu, A., \& Fritz, A. 2002, \mnras, 337, 256

\bibitem{} 
Balogh, M.~L., Morris, S.~L., Yee, H.~K.~C., Carlberg, R.~G., \& Ellingson,
E. 1999, \apj, 527, 54

\bibitem{} 
Balogh, M.~L., Navarro, J.~F., \& Morris, S.~L. 2000, \apj, 540, 113

\bibitem{} 
Barger, A. J., et al. 1998, \apj, 501, 522

\bibitem{} 
Barger, A.~J., Arag\'on-Salamanca, A., Ellis, R.~S., Couch, W.~J., Smail, I.,
\& Sharples, R.~M. 1996, \mnras, 279, 1

\bibitem{} 
Barnes, J. E. 2002, \mnras, 333, 481

\bibitem{} 
Barnes, J. E., \& Hernquist, L. E. 1991, \apj, 370, L65

\bibitem{} 
Barrientos, L., Manterola, M. C., Gladders, M. D., Yee, H. K. C., Infante, L., 
Hall, P., \& Ellingson, E. 2003, in Carnegie Observatories Astrophysics Series, 
Vol. 3: Clusters of Galaxies: Probes of Cosmological Structure and Galaxy 
Evolution, ed. J. S. Mulchaey, A. Dressler, \& A. Oemler (Pasadena: Carnegie 
Observatories, 
http://www.ociw.edu/ociw/symposia/series/symposium3/proceedings.html)

\bibitem{} 
Barton, E. J., Geller, M. J., \& Kenyon, S. J., 2000, \apj, 530, 660

\bibitem{} 
Benson, A.~J., Bower, R.~G., Frenk, C.~S., \& White, S.~D.~M. 2000, \mnras, 
314, 557

\bibitem{} 
Bond, J. R., Kaiser, N., Efstathiou, G., \& Cole, S. 1991, \apj, 379, 440

\bibitem{} 
Bower, R. G. 1991, \mnras, 248, 332

\bibitem{} 
Bower, R. G., Kodama, T., \& Terlevich, A. 1998, \mnras, 299, 1193

\bibitem{} 
Bower, R. G., Lucey, J. R., \& Ellis, R. S. 1992, \mnras, 254, 601

\bibitem{} 
Butcher, H., \& Oemler, A., Jr. 1978, \apj, 219, 18

\bibitem{} 
------. 1984, \apj, 285, 426

\bibitem{} 
Carlberg, R.~G., Yee, H.~K.~C., Morris, S.~L., Lin, H., Hall, P.~B.,
Patton, D.~R., Sawicki, M., \& Shepherd, C.~W. 2001, \apj, 563, 736

\bibitem{} 
Cohen, J.~G., Hogg, D.~W., Blandford, R.~D., Cowie, L.~L., Hu, E., Songalia,
A., Shopbell, P., \& Richberg, K. 2000, \apj, 538, 29

\bibitem{cole00}
Cole, S., Lacey, C. G., Baugh, C. M., \& Frenk, C. S. 2000, \mnras, 319, 168

\bibitem{} 
Couch, W. J., Balogh, M. L., Bower, R. G., Smail, I., Glazebrook, K., \& 
Taylor, M. 2001, \apj, 549, 820

\bibitem{} 
Couch, W. J., \& Newell, E. B. 1984, \apjs, 56, 143

\bibitem{} 
Couch, W. J., \& Sharples, R. M. 1987, \mnras, 229, 423

\bibitem{} 
Demarco, R., Rosati, P., Lidman, C., Nonino, M., Mainieri, V., Stanford, A., 
Holden, B., \& Eisenhardt, P. 2003, in Carnegie Observatories Astrophysics 
Series, Vol. 3: Clusters of Galaxies: Probes of Cosmological Structure and 
Galaxy Evolution, ed. J. S. Mulchaey, A. Dressler, \& A. Oemler (Pasadena: 
Carnegie Observatories, 
http://www.ociw.edu/ociw/symposia/series/symposium3/proceedings.html)

\bibitem{} 
De Propris, R., Stanford, S. A., Eisenhardt, P. R., Dickinson, M., \& Elston, 
R. 1999, \apj, 118, 719

\bibitem{} 
Diaferio, A., Kauffmann, G., Balogh, M. L., White, S. D. M., Schade, D., \& 
Ellingson, E. 2001, \mnras, 323, 999

\bibitem{} 
Dressler, A. 1980, \apjs, 42, 565

\bibitem{} 
Dressler, A., et al. 1997, \apj, 490, 577

\bibitem{} 
Dressler, A., \& Gunn, J. E. 1992, \apjs, 78, 1

\bibitem{} 
Drinkwater, M. J., Gregg, M. D., Holman, B. A., \& Brown, M. J. I. 2001, 
\mnras, 326, 1076

\bibitem{} 
Duc, P.-A., et al. 2002, \aa, 382, 60

\bibitem{} 
Eke, V., et al. 2003, in preparation

\bibitem{} 
Ellingson, E., Lin, H., Yee, H. K. C., \& Carlberg, R. G., 2001, \apj, 547, 609

\bibitem{} 
Ellis, R.~S., Smail, I., Dressler, A., Couch, W.~J., Oemler, A., Jr.,
Butcher, H., \& Sharples, R.~M. 1997, \apj, 483, 582

\bibitem{} 
Fairley, B.~W., Jones, L.~R., Wake, D.~A., Collins, C.~A., Burke, D.~J.,
Nichol, R.~C., \& Romer, A.~K. 2002, \mnras, 330, 755

\bibitem{} 
Fang, T., Sembach, K. R., \& Canizares, C. R. 2003, \apj, 586, L49

\bibitem{} 
Faber, S. M., Trager, S. C., Gonz\'alez, J. J., \& Worthey, G. 1999, Ap\&SS, 
267, 273

\bibitem{} 
Ferreras, I., Charlot, S. \& Silk, J. 1999, \apj, 521, 81

\bibitem{} 
Finn, R. A., \& Zaritsky, D. 2003, in Carnegie Observatories Astrophysics 
Series, Vol. 3: Clusters of Galaxies: Probes of Cosmological Structure and 
Galaxy Evolution, ed. J. S. Mulchaey, A. Dressler, \& A. Oemler (Pasadena: 
Carnegie Observatories,
 http://www.ociw.edu/ociw/symposia/series/symposium3/proceedings.html)

\bibitem{} 
Fritz, A., Ziegler, B. L., Bower, R. G., Smail, I., \& Davies, R. L. 2003, in 
Carnegie Observatories Astrophysics Series, Vol. 3: Clusters of Galaxies: 
Probes of Cosmological Structure and Galaxy Evolution, ed. J. S. Mulchaey, A. 
Dressler, \& A. Oemler (Pasadena: Carnegie Observatories, 
http://www.ociw.edu/ociw/symposia/series/symposium3/proceedings.html)

\bibitem{} 
Geller, M. J., \& Huchra, J. P. 1983, \apjs, 52, 61

\bibitem{} 
Gladders, M. D., Lopez-Cruz, O., Yee, H. K. C., \& Kodama, T. 1998, \apj, 
501, 571

\bibitem{} 
Glazebrook, K., Blake, C., Economou, F., Lilly, S., \& Colless, M. 1999, 
\mnras, 306, 843

\bibitem{} 
G\'omez, P. L., et al. 2003, \apj, 584, 210

\bibitem{} 
Goto, T., et al.  2003, PASJ, 55, 771

\bibitem{} 
Governato, F., Baugh, C. M., Frenk, C. S., Cole, S., Lacey, C. G., Quinn, T., 
\& Stadel, J. 1998, Nature, 392, 359

\bibitem{} 
Gunn, J. E., \& Gott, J. R. 1972, \apj, 176, 1

\bibitem{} 
Guzik, J., \& Seljak, U. 2002, \mnras, 335, 311

\bibitem{} 
G\'uzman, R., Lucey, J. R., Carter, D., \& Terlevich, R. J. 1992, \mnras, 
257, 187

\bibitem{} 
Helly, J. C., Cole, S., Frenk, C. S., Baugh, C. M., Benson, A., \& Lacey, C. 
2003, \mnras, 338, 903

\bibitem{} 
Henry, J. P., et al. 1995, \apj, 449, 422

\bibitem{} 
Hickson, P., Kindl, E., \& Auman, J. R. 1989, \apjs, 70, 687

\bibitem{} 
Hubble E., \& Humason, M. L. 1931, \apj, 74, 43

\bibitem{jenkins01}
Jenkins, A., Frenk, C. S., White, S. D. M., Colberg, J. M., Cole, S., Evrard, 
A. E., Couchman, H. M. P., \& Yoshida, N. 2001, \mnras, 321, 372

\bibitem{}
J{\o}rgensen, I. 1999, \mnras, 306, 607

\bibitem{}
J{\o}rgensen, I., Franx, M., Hjorth, J., \& van Dokkum, P. G. 1999, 
\mnras, 308, 833

\bibitem{} 
Kauffmann, G. 1995, \mnras, 274, 153

\bibitem{kauff99}
Kauffmann, G., Colberg, J. M., Diaferio, A., \& White, S. D. M. 1999, \mnras, 
303, 188

\bibitem{} 
Kelson, D. D., Illingworth, G. D., Franx, M., \& van~Dokkum, P. G. 2001, 
\apj, 552, L17

\bibitem{ken89} 
Kennicutt, R. C. 1989, \apj, 344, 685

\bibitem{} 
------. 1992, \apj, 388, 310

\bibitem{}
Kodama, T., \& Bower, R. G. 2001, \mnras, 321, 18

\bibitem{}
Kodama, T., Smail, I., Nakata, F., Okamura, S., \& Bower, R. G. 2001, \apj,
562, 9

\bibitem{} 
Lacey, C., \& Cole, S. 1993, \mnras, 262, 627

\bibitem{} 
Lambas, D. G., Tissera, P. B., Sol Alonso, M., \& Coldwell, G. 2003, \mnras, 
submitted (astro-ph/0212222)

\bibitem{} 
Larson, R. B., Tinsley, B. M., \& Caldwell, C. N. 1980, \apj, 237, 692

\bibitem{} 
Lewis, I. J., et al. 2002, \mnras, 334, 673

\bibitem{} 
Lilly, S. J., Tresse, L., Hammer, F., Crampton, D., \& Le F\'evre, O. 1995, 
\apj, 455, 108

\bibitem{} 
Madau, P., Pozzetti, L., \& Dickinson, M. 1998, \apj, 498, 106

\bibitem{} 
Margoniner, V. E., de Carvalho, R. R., Gal, R. R., \& Djorgovski, S. G. 2001, 
\apj, 548, L143

\bibitem{martin99}
Martin, C. L. 1999, \apj, 513, 156

\bibitem{} 
McIntosh, D. H., Rix, H.-W., \& Caldwell, N. 2003, \apj, submitted 
(astro-ph/0212427)

\bibitem{} 
Metavier, A. J. 2003, in Carnegie Observatories Astrophysics Series, Vol. 3: 
Clusters of Galaxies: Probes of Cosmological Structure and Galaxy Evolution, 
ed. J. S. Mulchaey, A. Dressler, \& A. Oemler (Pasadena: Carnegie Observatories,
http://www.ociw.edu/ociw/symposia/series/symposium3/proceedings.html)

\bibitem{} 
Mihos, J. C. 2003, in Carnegie Observatories Astrophysics Series, Vol. 3: 
Clusters of Galaxies: Probes of Cosmological Structure and Galaxy Evolution, 
ed. J. S. Mulchaey, A. Dressler, \& A. Oemler (Cambridge: Cambridge Univ.\ 
Press), in press

\bibitem{} 
Miller, N. A. 2003, in Carnegie Observatories Astrophysics Series, Vol. 3: 
Clusters of Galaxies: Probes of Cosmological Structure and Galaxy Evolution, 
ed. J. S. Mulchaey, A. Dressler, \& A. Oemler (Pasadena: Carnegie 
Observatories, 
http://www.ociw.edu/ociw/symposia/series/symposium3/proceedings.html)

\bibitem{} 
Miller, N. A., \& Owen, F. N. 2002, \aj, 124, 2453

\bibitem{} 
Milvang-Jensen, B., Arag\'on-Salamanca, A., Hau, G., J{\o}rgensen, I., \& 
Hjorth, J. 2003, \mnras, 339, 1

\bibitem{} 
Mo, H. J., \& White, S. D. M. 2002, \mnras, 336, 112

\bibitem{} 
Moore, B., Frenk, C. S., \& White, S. D. M. 1993, \mnras, 261, 827

\bibitem{} 
Moore, B., Katz, N., Lake, G., Dressler, A., \& Oemler, A. 1996, Nature, 379, 
613

\bibitem{} 
Morrison, G. E., \& Owen, F. N. 2003, \aj, 125, 506

\bibitem{} 
Mulchaey, J. S., Davis, D. S., Mushotzky, R. F., \& Burstein, D. 2003, \apjs, 
145, 39

\bibitem{} 
Nakata, F., et al. 2003, in preparation

\bibitem{} 
Oemler, A., Jr. 1974, \apj, 194, 10

\bibitem{} 
Poggianti, B. M. 2003, in Carnegie Observatories Astrophysics Series, Vol. 3: 
Clusters of Galaxies: Probes of Cosmological Structure and Galaxy Evolution,
ed. J. S. Mulchaey, A. Dressler, \& A. Oemler (Cambridge: Cambridge Univ.\ 
Press), in press

\bibitem{} 
Poggianti, B. M., et al. 2001, \apj, 563, 118

\bibitem{} 
Poggianti, B.~M., Smail, I., Dressler, A., Couch, W.~J., Barger, A.~J.,
Butcher, H., Ellis, R.~S., \& Oemler, A., Jr. 1999, \apj, 518, 576

\bibitem{} 
Poggianti, B. M., \& Wu, H. 2000, \apj, 529, 157 

\bibitem{} 
Press, W. H., \& Schechter, P. 1974, \apj, 187, 425

\bibitem{} 
Quilis, V., Moore, B., \& Bower, R. G. 2000, Science, 288, 1617

\bibitem{} 
Quintero, A. D., et al. 2003, \apj, submitted (astro-ph/0307074)

\bibitem{} 
Rakos, K. D., \& Schombert, J. M. 1995, \apj, 439, 47

\bibitem{} 
Rudnick, G. H., De Lucia, G., White, S. D. M., \& Pell\'o, R. 2003, in 
Carnegie Observatories Astrophysics Series, Vol. 3: Clusters of Galaxies:
Probes of Cosmological Structure and Galaxy Evolution, ed. J. S. Mulchaey,
A. Dressler, \& A. Oemler (Pasadena: Carnegie Observatories,
http://www.ociw.edu/ociw/symposia/series/symposium3/proceedings.html)

\bibitem{} 
Sakamoto, T., Chiba, M., \& Beers, T. C. 2003, \aa, 397, 899

\bibitem{} 
Sandage A., 1961, in The Hubble Atlas of Galaxies (Washington, DC: Carnegie 
Int. of Washington)

\bibitem{} 
Sandage A., \& Visvanathan, N. 1978, \apj, 223, 707

\bibitem{schm59} 
Schmidt, M. 1959, \apj, 129, 243

\bibitem{} 
Sheth, R. K., Mo, H. J., \& Tormen, G. 2001, \mnras, 323, 1

\bibitem{} 
Smail, I., Edge, A. C., Ellis, R. S., \& Blandford, R. D. 1998, \mnras, 293, 124

\bibitem{} 
Smail, I., Morrison, G., Gray, M. E., Owen, F. N., Ivison, R. J., Kneib, J. P.,
\& Ellis, R. S. 1999, \apj, 525, 609

\bibitem{} 
Somerville, R. S., \& Primack, J. R. 1999, \mnras, 310, 1087

\bibitem{} 
Springel, V., White, S. D. M., Tormen, G., \& Kauffmann, G. 2001, \mnras, 328, 
726

\bibitem{} 
Tran, K., Franx, M., Illingworth, G., \& van Dokkum, P. 2003, in
Carnegie Observatories Astrophysics Series, Vol. 3: Clusters of Galaxies:
Probes of Cosmological Structure and Galaxy Evolution, ed. J. S. Mulchaey,
A. Dressler, \& A. Oemler (Pasadena: Carnegie Observatories,
http://www.ociw.edu/ociw/symposia/series/symposium3/proceedings.html)

\bibitem{} 
Treu, T. 2003, in Carnegie Observatories Astrophysics Series, Vol. 3:
Clusters of Galaxies: Probes of Cosmological Structure and Galaxy Evolution,
ed. J. S. Mulchaey, A. Dressler, \& A. Oemler (Cambridge: Cambridge Univ.\
Press), in press

\bibitem{} 
Treu, T., Ellis, R. S., Kneib, J.-P., Dressler, A., Smail, I., Czoske, O., 
Oemler, A., \& Natarajan P. 2003, \apj, 591, 53

\bibitem{} 
van Dokkum, P. G., \& Franx, M. 2001, \apj, 553, 90

\bibitem{} 
van Dokkum, P. G., Franx, M., Fabricant, D., Illingworth, G. D., \& Kelson, 
D. D. 2000, \apj, 541, 95

\bibitem{} 
van Dokkum, P. G., Franx, M., Kelson, D. D., Illingworth, G. D., Fischer, D., 
\& Fabricant, D. 1998, \apj, 500, 714

\bibitem{} 
van Dokkum, P. G., \& Stanford, S. A. 2003, \apj, 585, 78

\bibitem{} 
Wilson, G., Cowie, L. L., Barger, A. J., \& Burke, D. J. 2002, \aj, 124, 1258

\bibitem{} 
Zabludoff, A. I., Zaritsky, D., Lin, H., Tucker, D., Hashimoto, Y., Shectman, 
S. A., Oemler, A., \& Kirshner, R. P. 1996, \apj, 466, 104

\bibitem{} 
Ziegler, B., B\"ohm, A., Jager, K., Fritz, A., \& Heidt J. 2003, in Carnegie 
Observatories Astrophysics Series, Vol. 3: Clusters of Galaxies: Probes of 
Cosmological Structure and Galaxy Evolution, ed. J. S. Mulchaey, A. Dressler, 
\& A. Oemler (Pasadena: Carnegie Observatories,
http://www.ociw.edu/ociw/symposia/series/symposium3/proceedings.html)

\end{thereferences}
\end{document}